\documentclass[aps,twocolumn,showpacs,prl]{revtex4-1}
\usepackage{graphicx,graphics}
\usepackage{amsmath,amssymb,amsfonts}
\usepackage{multirow}
\usepackage{hyperref}

\begin{document}
\title{Fractional dynamics on networks: Emergence of anomalous diffusion and L\'evy flights}
\author{A. P. Riascos}
\author{Jos\'e L. Mateos}
\affiliation{Instituto de F\'isica, Universidad Nacional Aut\'onoma de M\'exico, 
Apartado Postal 20-364, 01000 M\'exico, D.F., M\'exico}
\date{\today}

\begin{abstract}
We introduce a formalism of fractional diffusion on networks based on a fractional Laplacian matrix that can be constructed directly from the eigenvalues and eigenvectors of the Laplacian matrix. This fractional approach allows random walks with long-range dynamics providing a general framework for anomalous diffusion and navigation, and inducing dynamically the small-world property on any network. We obtained exact results for the stationary probability distribution, the average fractional return probability and a global time, showing that the efficiency to navigate the network is greater if we use a fractional random walk in comparison to a normal random walk. For the case of a ring, we obtain exact analytical results showing that the fractional transition and return probabilities follow a long-range power-law decay, leading to the emergence of L\'evy flights on networks. Our general fractional diffusion formalism applies to regular, random and complex networks and can be implemented from the spectral properties of the Laplacian matrix, providing an important tool to analyze anomalous diffusion on networks.            
\\[1mm]
\end{abstract}

\pacs{89.75.Hc, 05.40.Fb, 02.50.-r, 05.60.Cd}

\maketitle
\section{Introduction}
The recent burst of work around the idea of networks can be explained by the importance of this concept in a vast range of fields, which includes both the structural features and the functional properties of networks \cite{NewmanBook,Bocca,ArenasPhysRep2008,VespiBook,VespiNature2012}. In particular, we are interested in random walks taking place on networks, such as regular \cite{Hughes,Weiss,KlafterSokolov}, random and complex networks \cite{NohRieger}, and more recently temporal \cite{StarniniPRE2012,PerraPRL2012,GauvinSciRep2013}, multiplex \cite{GomezPRL2013,SolePRE2013,DomenicoPRX2013}, and interconnected networks \cite{RadiPRX2014,DomenicoPNAS2014}. Random walks are useful to analyze problems of searching and navigability on networks, with applications to many different fields, such as the propagation of epidemics, traffic flow, and rumor and information spreading \cite{VespiBook,VespiNature2012}; for a recent survey, see Ref. \cite{HuangPhysA2014}.

In a recent paper, the usual navigation strategy of transitions to nearest neighbors is generalized by allowing long-range navigation on complex networks using L\'evy random walks \cite{RiascosMateos}. This generalization allows transitions not only to nearest neighbors but to second-, third- or $n$-nearest neighbors. This new strategy was inspired by the study of L\'evy flights, where the lengths of random displacements obey asymptotically a power-law probability distribution \cite{MetzlerPhysRep2000}. These L\'evy flights generate anomalous diffusion \cite{BouchaudPhysRep1990} and have been used as searching and navigation strategies by animals \cite{BES2004,BoyerPRS2006,SimsNature2008,SimsNature2010,JagerScience2011,ViswaBook2011,FedeBook2014} and also in human mobility and behavior \cite{Brock2006,Gonzalez2008,Song2010,Rhee2011,Simini2012,RadiPlosONE2012,RadiPRE2012,RadiCSF2013,FedeBook2014,PNAS2014}.

On the other hand, it is well known that one can study anomalous diffusion, and in particular L\'evy flights, 
using a fractional calculus approach \cite{MetzlerPhysRep2000,Restaurant,KlafterSokolov}. Likewise, we introduce 
in this paper a fractional approach applied directly to the dynamics on networks to address the problem of 
anomalous diffusion and long-range navigation. Our fractional analysis applied to general networks provides us 
with a framework to deal with a richer dynamics on complex networks that includes, among other things, 
L\'evy flights \cite{RiascosMateos}.

%
%
\begin{figure}[!b] 
\begin{center}
\includegraphics*[width=0.36\textwidth]{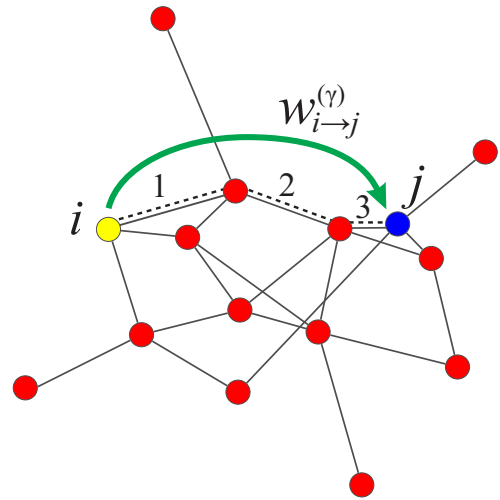}
\end{center}
\vspace{-3mm}
\caption{\label{Fig1}(Color online) A random walker performing long-range transitions on a network can move directly from node $i$ to 
node $j$ with a transition probability given by $w^{(\gamma)}_{i\to j}$ (the index $\gamma$ refers to a fractional 
dynamics described in the text). The nodes are three degrees apart, that is, the geodesic distance is three, as indicated 
by the dashed line. The geodesic distance is the integer number of steps of the shortest path connecting two nodes. 
Using this long-range dynamics one can contact directly long-distance nodes without the intervention of intermediate 
nodes and without altering the topology of the network.} 
\end{figure}

This generalized long-range navigation can consider some common situations in real networks. For instance, 
in social networks one can use the knowledge of the network beyond our first direct contacts or acquaintances. 
Currently, using social networks sites, one can identify the friends of your acquaintances (second-nearest neighbors) 
or the friends of the friends of your acquaintances (third-nearest neighbors) and so forth to search for information, 
a job, an expert opinion, etc. In this fashion, one can contact people two or three degrees away of your friend 
directly, without the intervention of your friend. This situation, which we all use almost on a daily basis, 
corresponds to a long-range navigation on a network: a social network in this case. We illustrate this dynamics 
in Fig. \ref{Fig1}, where we depict a general network of nodes and links among them. A random walker performing 
long-range transitions can move directly from node $i$ to node $j$ with a transition probability given by 
$w^{(\gamma)}_{i\to j}$ (the index $\gamma$ refers to a fractional dynamics described below). The nodes are three degrees 
apart, that is, the geodesic distance is three, as indicated by the dashed line. The geodesic distance is the 
integer number of steps of the shortest path connecting two nodes. In this long-range fractional dynamics on 
networks one can contact directly long-distance nodes without the intervention of intermediate nodes and without 
altering the topology of the network.

We start with an overview of the formalism associated with diffusion processes and normal random walks on networks. 
Then, we extend the formalism to the case of fractional diffusion on networks and obtain a random walker defined by 
a transition probability matrix that allows a long-range dynamics.  We deduce the stationary probability 
distribution of this process and, in order to study the efficiency of the random walker, we calculate the average 
return probability to a node and the average global time associated to the capacity to explore the network. 
\section{Dynamics on networks}
We consider undirected connected networks with $N$ nodes, described by the adjacency matrix $\mathbf{A}$ with elements $A_{ij}=A_{ji}=1$ (where $i,j=1,\ldots ,N$) if there is a link between node $i$ and node $j$, and $A_{ij}=A_{ji}=0$ otherwise. We take $A_{ii}=0$ to avoid loops in the network. The degree of the node $i$ is given by $k_i=\sum_{l=1}^N A_{il}$. The Laplacian matrix $\mathbf{L}$ is defined as $L_{ij}=\delta_{ij} k_i-A_{ij}$, where $\delta_{ij}$ denotes the Kronecker delta; this matrix $\mathbf{L}$ is interpreted as a discrete version of the operator $(-\nabla^2)$ \cite{NewmanBook}. Hence, the diffusion equation in a network takes the form \cite{NewmanBook,VespiBook,Weiss}:
\begin{equation}\label{DiffGraph}
	\mathbf{L}\left|\psi(t)\right\rangle=-\frac{d \,}{dt}\left|\psi(t)\right\rangle.
\end{equation}
The vector $\left|\psi(t)\right\rangle$ describes the system at time $t$, $\left|\psi(t)\right\rangle=\sum_{m=1}^N a_m(t)\left|m\right\rangle$, where $\{\left|m\right\rangle\}_{i=1}^N$ represents the canonical base of $\mathbb{R}^N$ \cite{Mulken}. On the  other hand, random walks on networks are defined in terms of the modified Laplacian $\mathcal{L}$ with elements $\mathcal{L}_{ij}=\frac{L_{ij}}{L_{ii}}=\delta_{ij}-w_{i\to j}$, where $w_{i\to j}=\frac{A_{ij}}{k_i}$ are the elements of the transition matrix $\mathbf{W}$ of the normal random walk on a network, describing transitions only to nearest neighbors with equal probability \cite{Hughes,NohRieger}. For continuous time, the dynamics of the random walker is determined by the master equation \cite{VespiBook}:
\begin{equation} \label{master}
\frac{d p_{ij}(t)}{dt}=-\sum_{l=1}^N\mathcal{L}_{lj}\, p_{il}(t), 
\end{equation}
where $p_{ij}(t)$ denotes the probability  to find the random walker in the node $j$ at time $t$ starting from the node $i$ at $t=0$. The master equation, Eq. (\ref{master}), describes a Markovian process with a stationary distribution $p_j^{\infty}=\lim_{t \to \infty}p_{ij}(t)$, i.e. the probability for a walker to be in node $j$ in the limit of large times. For a normal random walk is given by $p_j^\infty =\frac{k_j}{\sum_{m=1}^N k_m}$ \cite{VespiBook,NohRieger}. Another important quantity in the study of the diffusive transport is the average return probability defined by $p_0(t)=\frac{1}{N}\sum_{i=1}^N p_{ii}(t)$ \cite{VespiBook,Blumen,Mulken}. From Eq.  (\ref{master}) it can be shown that $p_0(t)=\frac{1}{N}\sum_{m=1}^N \exp[-\zeta_m t]$, where $\{\zeta_i\}_{i=1}^N$  are the eigenvalues of the modified Laplacian $\mathcal{L}$ \cite{VespiBook}  .
\section{Fractional dynamics on networks}
Having defined the Laplacian matrix $\mathbf{L}$ and the modified Laplacian matrix $\mathcal{L}$ related to normal random walks, we introduce a generalization of these concepts in order to study the fractional diffusion on networks. For recent reviews of the fractional calculus approach to anomalous diffusion, see Refs. \cite{MetzlerPhysRep2000,Restaurant,KlafterSokolov}.
\\
We are interested in studying a generalization of Eq. (\ref{DiffGraph}) that reads:
\begin{equation}
	\mathbf{L}^{\gamma}\left|\psi(t)\right\rangle=-\frac{d \,}{dt}\left|\psi(t) \right\rangle \qquad  0<\gamma< 1 \, ,
\end{equation}
where $\mathbf{L}^{\gamma}$ is the Laplacian matrix $\mathbf{L}$ to a power $\gamma$, where $\gamma$ is a real number ($0<\gamma<1$). 
\\
In the limit where $\gamma\to 1$, we recover Eq. (\ref{DiffGraph}). In the following part we study the consequences of this definition and the characteristics of the random walks behind this dynamical process.

Since $\mathbf{L}$ is a symmetric matrix, using the Gram-Schmidt orthonormalization of the eigenvectors of $\mathbf{L}$, we obtain a set of eigenvectors $\{\left|\Psi_j\right\rangle\}_{j=1}^N$  that satisfy the eigenvalue equation 
$\mathbf{L}\left|\Psi_j\right\rangle=\mu_j\left|\Psi_j\right\rangle$ for $j=1,\ldots,N$ 
and $\left\langle\Psi_i|\Psi_j\right\rangle=\delta_{ij}$, where $\mu_j$ are the eigenvalues, which are real and nonnegative. For connected networks, the smallest eigenvalue $\mu_1=0$ and $0 < \mu_m$ for $m=2,\ldots,N$ \cite{VanMieghem}. We define the orthonormal matrix $\mathbf{Q}$  with elements $Q_{ij}=\left\langle i|\Psi_j\right\rangle$  and the diagonal matrix $\mathbf{\Lambda}=\text{diag}(0,\mu_2,\ldots,\mu_N)$. These matrices satisfy $\mathbf{L}\,\mathbf{Q}=\mathbf{Q}\,\mathbf{\Lambda}$, therefore $\mathbf{L}=\mathbf{Q}\mathbf{\Lambda}\mathbf{Q}^{T}$, where $\mathbf{Q}^T$ denotes the transpose of $\mathbf{Q}$. Therefore, we have \cite{bellman1960}:
\begin{equation}\label{LfracDef}
	\mathbf{L}^{\gamma}=\mathbf{Q} \mathbf{\Lambda}^{\gamma} \mathbf{Q}^{T}
	=\sum_{m=2}^N \mu_m^{\gamma}\left|\Psi_m\right\rangle\left\langle \Psi_m\right| ,
\end{equation}
where $\mathbf{\Lambda}^{\gamma}=\text{diag}(0,\mu_2^{\gamma},\ldots,\mu_N^{\gamma})$. In this way, Eq. (\ref{LfracDef}) gives the spectral form of the fractional Laplacian matrix, and therefore,
\begin{equation}
\mathbf{L}^\gamma\left|\Psi_j\right\rangle=\mu_j^\gamma\left|\Psi_j\right\rangle \qquad \text{for}\quad j=1,\ldots,N.
\end{equation}
This result indicates that in order to treat the fractional dynamics we can simply calculate the spectrum $\{\mu_j\}_{j=1}^N$ and then calculate $\{\mu_j^\gamma\}_{j=1}^N$. 

On the other hand, in analogy with the matrix $\mathcal{L}$, we define the modified fractional Laplacian matrix $\mathcal{L}^{(\gamma)}$ with elements $\mathcal{L}^{(\gamma)}_{ij}=(\mathbf{L}^{\gamma})_{ij}/(\mathbf{L}^{\gamma})_{ii}$. This modified fractional Laplacian is related to the dynamics of a random walker on a network determined by a fractional transition matrix $\mathbf{W}^{(\gamma)}$ with elements $w^{(\gamma)}_{i\to j}$ given by $w^{(\gamma)}_{i\to j} = \delta_{ij} -\mathcal{L}^{(\gamma)}_{ij}$. Therefore: 
\begin{equation}\label{wfrac}
w_{i\rightarrow j}^{(\gamma)}=\delta_{ij}	-\frac{(\mathbf{L}^{\gamma})_{ij}}{k_i^{(\gamma)}} \, ,
\end{equation}
where we define the \textit{fractional degree} of the node $i$ as $k_i^{(\gamma)}\equiv(\mathbf{L}^{\gamma})_{ii}$. Notice that $w_{i\rightarrow i}^{(\gamma)}=0$. Also, the fractional transition matrix for $0<\gamma\leq 1$ is a stochastic matrix that satisfies $\sum_{l=1}^{N} w_{i\to l}^{(\gamma)}=1$. On the other hand, from Eq. (\ref{wfrac}) in the case $\gamma=1$, we recover the normal random walk with a transition matrix given by $w_{i\rightarrow j}^{(1)}=A_{ij}/k_i$.
\\

Now, the corresponding fractional stationary distribution $p_{i}^{\infty} (\gamma)$ of the random walker, from Eq. (\ref{wfrac}), and using the detailed balance condition $k_i^{(\gamma)}p_j^{\infty} (\gamma) = k_j^{(\gamma)}p_i^{\infty}(\gamma) $ \cite{RiascosMateos}, is given by
\begin{equation}\label{PinfFrac}
p_i^{\infty} (\gamma) =\frac{(\mathbf{L}^{\gamma})_{ii}}{\sum_{m=1}^N (\mathbf{L}^{\gamma})_{mm}} =\frac{k^{(\gamma)}_i}{\sum_{m=1}^N k^{(\gamma)}_m}.
\end{equation}
This is a generalization of the result $p_i^{\infty}\propto k_i$ for normal random walks discussed before, and is recovered from Eq. (\ref{PinfFrac}) when $\gamma=1$. The general result that relates this stationary distribution with the mean first return time still applies and reads: $\langle T_{ii} ^{(\gamma)} \rangle = 1/p_i^{\infty} (\gamma) $ \cite{RiascosMateos}. 
%
%
%
%
%
\begin{figure*}[!t] 
\begin{center}
\includegraphics*[width=0.95\textwidth]{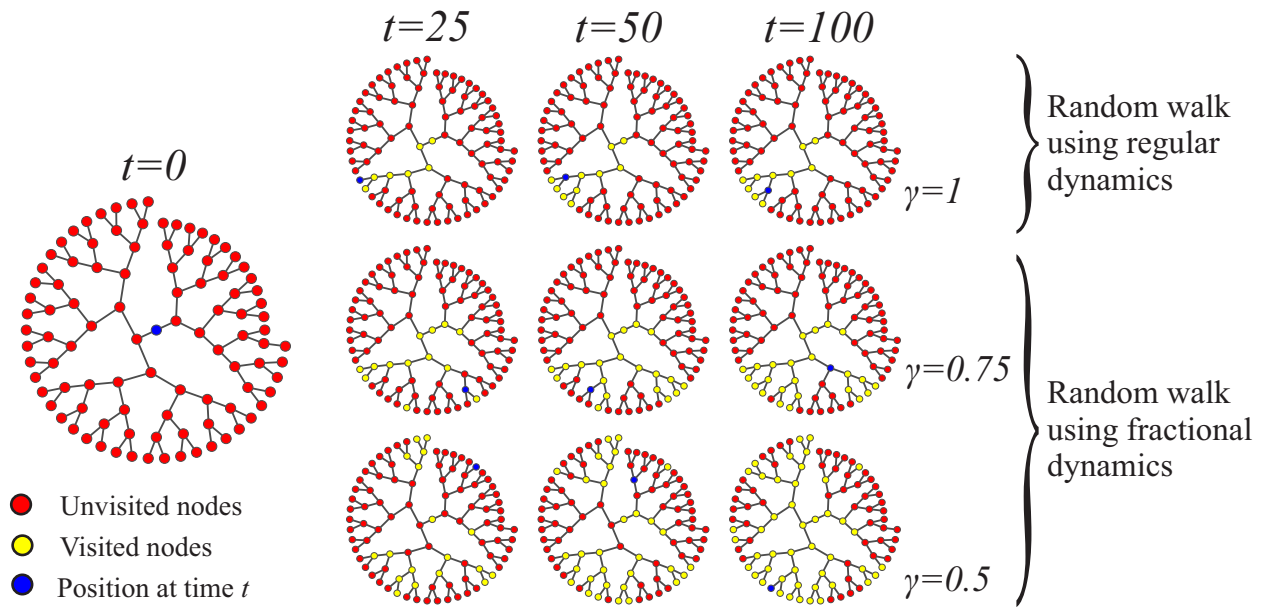}
\end{center}
\vspace{-5mm}
\caption{\label{Fig2}(Color online) Monte Carlo simulation of a discrete-time random walker on a 
network. We choose for clarity a tree, with $N=100$ nodes, but the qualitative results stand for any
network. The discrete time $t$ denotes the number of steps of the random walker as it moves from one node to the next node
on the network. This discrete random process is governed by a master equation with a transition probability matrix that
gives the probability of moving from node to node. The dynamics starts at $t = 0$ from an arbitrary node. 
We show three discrete times $t=25,50,100$ for three values of the
parameter $\gamma=1,0.75,0.5$. Here, we depict one representative realization of a random walker as it navigates 
from one node to another randomly. The case $\gamma=1$ corresponds to normal random walk leading to normal diffusion. In 
this case, the random walker can move only locally to nearest neighbors and, as can be seen in the figure, the walker
revisits very frequently the same nodes and therefore the exploration of the network is redundant and not very 
efficient. The cases with $\gamma=0.75,0.5$ correspond to a fractional random walk leading to anomalous diffusion.
In this case, the random walker can navigate in a long-range fashion from one 
node to another arbitrarily distant node. This allows to explore more efficiently the network since the walker does not
tend to revisit the same nodes; on the contrary, it tends to explore and navigates distant new regions each time. 
All this can be seen in the figure for different times, and allow us to make a comparison between a random walker 
using regular dynamics and fractional dynamics.} 
\end{figure*}
\\

The modified Laplacian $\mathcal{L}$ describes a random walker using a transition matrix that allows only the passage from a node to one of its neighbors (which we recover when $\gamma=1$). In what follows we show that in the fractional case, when $0<\gamma<1$, the random walker can move using long-range navigation on the network. 
%
For this purpose, we start with the example of a ring (1D lattice with periodic boundary conditions), where the 
eigenvalues of the Laplacian matrix are $\mu_m=2-2\cos[2\pi(m-1)/N]$  and the corresponding eigenvectors have 
components $\langle m | \Psi_l \rangle=e^{-\text{i}\frac{2\pi}{N}(l-1)(m-1)}/\sqrt{N}$ 
(where $\text{i}\equiv\sqrt{-1}$) \cite{VanMieghem}. Using Eq. (\ref{LfracDef}) we obtain:
\begin{align}\label{LfracRingijExp}
(\mathbf{L}^{\gamma})_{ij}&=\frac{1}{N}\sum_{l=2}^N \left(2-2\cos\phi_l\right)^\gamma e^{-\text{i}\phi_l(i-j)}\\
\label{LfracRingij}
&=\frac{1}{N}\sum_{l=2}^N\left(2-2\cos\phi_l\right)^\gamma\cos\left[\phi_l(i-j)\right]
\end{align}
with $\phi_m=\frac{2\pi}{N}(m-1)$. In Eq. (\ref{LfracRingijExp}), the imaginary part associated 
with $\sin\left[\phi_l(i-j)\right]$ cancels out and only the real part remains in Eq. (\ref{LfracRingij}). 
Now we present this result in terms of the geodesic distance $d_{ij}$, defined as the integer number of steps 
of the shortest path connecting node $i$ and node $j$. For the case of a ring:
\begin{equation*}
d_{ij}=
\begin{cases}
\left|i-j\right| \quad &\text{if} \quad \left|i-j\right|=0,1,\ldots,\left\lfloor \frac{N}{2}\right\rfloor,\\
N-\left|i-j\right|  &\text{if} \quad \left|i-j\right|=\left\lfloor \frac{N}{2}\right\rfloor+1,\ldots, N,\\
\end{cases}
\end{equation*}
where $\left\lfloor x \right\rfloor$ is the floor function that gives the largest integer not greater than $x$. 
The distances $d_{ij}$ for a ring satisfy 
\begin{equation}
\cos\left[\frac{2\pi}{N}(l-1)(i-j)\right]= \cos\left[\frac{2\pi}{N}(l-1)d_{ij}\right],
\end{equation}
and using this result in Eq. (\ref{LfracRingij}), we have for the fractional Laplacian of a ring:
\begin{align}
\nonumber
(\mathbf{L}^{\gamma})_{ij}&=\frac{1}{N}\sum_{l=2}^N\left(2-2\cos\phi_l\right)^\gamma\cos\left[\phi_l d_{ij}\right]\\ 
\label{LfracCycle}
&=\frac{4^{\gamma}}{N}\sum_{l=2}^N\sin^{2\gamma}\left[\frac{\phi_l}{2}\right]\cos\left[\phi_l d_{ij}\right].
\end{align}
On the other hand, using the fact that $d_{ij}=0$ if $i=j$, we obtain from Eq. (\ref{LfracCycle}) the fractional degree:
\begin{equation}\label{KfracRing}
k_{i}^{(\gamma)}=(\mathbf{L}^{\gamma})_{ii}=\frac{4^{\gamma}}{N}\sum_{l=2}^N\sin^{2\gamma}\left[\frac{\phi_l}{2}\right].
\end{equation}
Now, introducing Eqs. (\ref{LfracCycle}) and (\ref{KfracRing})  in Eq. (\ref{wfrac}), we obtain the 
elements $w_{i\to j}^{(\gamma)}$ of the fractional transition matrix for a ring:
\begin{equation}\label{wijring}
w_{i\rightarrow j}^{(\gamma)}=\delta_{ij}-
\frac{\sum_{l=2}^N \sin^{2\gamma}\left[\frac{\pi(l-1)}{N}\right] \cos\left[\frac{2\pi(l-1)}{N}d_{ij}\right]}
{\sum_{l=2}^N \sin^{2\gamma}\left[\frac{\pi(l-1)}{N}\right]}\, .
\end{equation}
The result obtained in Eq. (\ref{wijring}) reveals the nonlocal character of the emergent process behind fractional diffusion on networks, where the transition probability depends explicitly on the global distance $d_{ij}$, at variance with the case of normal random walks, where the transition probability allows transitions only  to nearest neighbors.  In the limit $N\gg1$, the sum in Eq. (\ref{LfracRingijExp}) can be approximated by an integral  that takes the form:
\begin{equation}
(\mathbf{L}^{\gamma})_{ij}=\frac{1}{2\pi}\int_{0}^{2\pi} (2-2\cos\theta)^\gamma e^{\text{i}d_{ij}\theta}d\theta,
\end{equation}
which can be evaluated analytically and is given by (see Ref. \cite{Bounded} for a discussion):
\begin{equation}\label{LfracLine}
(\mathbf{L}^{\gamma})_{ij}=-\frac{\Gamma(d_{ij}-\gamma)\Gamma(2\gamma+1)}{\pi\Gamma(1+\gamma+d_{ij})}\sin(\pi \gamma),
\end{equation}
where $ \Gamma(x)$ is the Gamma function.
Using Eq. (\ref{LfracLine}) in Eq. (\ref{wfrac}) we obtain that for a ring:
\begin{equation}
w_{i\rightarrow j}^{(\gamma)}=\delta_{ij}-\frac{\Gamma(\gamma +1)\Gamma(d_{ij}-\gamma )}{\Gamma(-\gamma )\Gamma (d_{ij}+\gamma +1)}.
\end{equation}
For $i$, $j$ such that $d_{ij} \gg1$, and using the asymptotic result $\Gamma(n+\alpha)\approx\Gamma(n)n^\alpha$, for an integer $n\gg1$, and a real $\alpha$, we arrive at the result:
\begin{equation}
w_{i\rightarrow j}^{(\gamma)}\sim d_{ij}^{-(1+2\gamma)}.
\end{equation}
This asymptotic expression is not valid when $\gamma\to 0$ or $\gamma\to 1$. To summarize, for a very large ring and very large geodesic distances between nodes, we have shown explicitly that the transition probability that emerges from the fractional dynamics is a power law that corresponds precisely to the L\'evy random walks introduced in Ref. \cite{RiascosMateos}.  
\\  
%
%
%

This long-range navigation, based on power laws of the shortest paths introduced by Ref. \cite{RiascosMateos} is valid 
for general networks and has been explored by other authors afterwards \cite{LinPRE2013,HuangPhysA2014,ZhaoPhysA2014}. 
For this long-range fractional dynamics, the transition probability matrix can connect arbitrarily distant nodes and, 
in this sense, the problem can be mathematically equivalent to an abstract complete weighted network.
\\   
 
In order to illustrate the effect of the fractional dynamics of a random walker on a network,  
in Fig. \ref{Fig2} we show the Monte Carlo simulation of a discrete-time random walker on a 
network. We choose for clarity a tree (network without loops \cite{NewmanBook}) but the qualitative results stand for any
network. The discrete time $t$ denotes the number of steps of the random walker as it moves from one node to the next node
on the network. This discrete random process is governed by a master equation with a transition probability matrix that
gives the probability of moving from node to node. Given the topology of the network, we calculate the adjacency matrix 
and the corresponding Laplacian matrix $\mathbf{L}$ of the network. Then we obtain its eigenvalues and eigenvectors 
that allow us in turn to get the fractional Laplacian matrix $\mathbf{L}^{\gamma}$. Finally, using Eq. (\ref{wfrac}), 
we determine the transition probabilities for different values of the parameter $\gamma$.
The dynamics starts at $t = 0$ from an arbitrary node. We show three discrete times $t=25,50,100$ for three values of the
parameter $\gamma=1, 0.75, 0.5$. Here, we depict one representative realization of a random walker as it navigates 
from one node to another randomly. The case $\gamma=1$ corresponds to normal random walk leading to normal diffusion. In 
this case, the random walker can move only locally to nearest neighbors and, as can be seen in the figure, the walker
revisits very frequently the same nodes and therefore the exploration of the network is redundant and not very 
efficient. The cases with $\gamma = 0.75, 0.5$ correspond to a fractional random walk leading to anomalous diffusion.
In this case, the random walker can navigate in a long-range fashion from one 
node to another arbitrarily distant node. This allows us to explore more efficiently the network since the walker does not
tend to revisit the same nodes; on the contrary, it tends to explore and navigates distant new regions each time. 
All this can be seen in the figure for different times, and allow us to make a comparison between a random walker 
using regular dynamics and a fractional dynamics; see the Supplementary 
Material \footnote{See Supplemental Material for videos with the Monte Carlo simulation of the 
random walker. The cases with $\gamma=1,0.75,0.5$ and $0.25$ are presented in the videos video1.avi, video2.avi, video3.avi, and video4.avi, respectively.}. 

Finally, it is important to stress that this general dynamics given by the fractional Laplacian matrix $\mathbf{L}^{\gamma}$ has embedded the seeds of a long-range dynamics on networks of any kind. Not only involving the shortest paths, but also trajectories of any kind in the network.  That is, the fractional transition probabilities introduced in Eq. (\ref{wfrac}) contains a global dynamics in networks. Here lies its importance and the fruitful applications that it may have for many real processes in networks. 
\section{Fractional return probability}
Now, we analyze the continuous-time random walks defined by the fractional transition matrix $\mathbf{W}^{(\gamma)}$. Using the fractional matrix $\mathcal{L}^{(\gamma)}$  in the master equation (\ref{master}), the fractional occupation probability $p_{ij}^{(\gamma)} (t)$ satisfies:
\begin{equation}\label{masterfrac}
\frac{d p_{ij}^{(\gamma)} (t)}{dt} =-\sum_{l=1}^N\mathcal{L}_{lj}^{(\gamma)}\, p_{il}^{(\gamma)} (t). 
\end{equation}
We are interested in the efficiency of the dynamics, described by Eq. (\ref{masterfrac}), to explore the network. In the fractional case a treatment similar to the analysis of Eq. (\ref{master}) allows us to obtain the average fractional return probability, defined by $p_0^{(\gamma)}(t)=\frac{1}{N}\sum_{i=1}^N p_{ii}^{(\gamma)}(t)$, as:
\begin{equation}\label{Pretfrac}
p_0^{(\gamma)} (t)=\frac{1}{N}\sum_{i=1}^N \exp[-\zeta_i^{(\gamma)}t]  ,               
\end{equation}
where $\zeta^{(\gamma)}_1=0$, $0 < \zeta^{(\gamma)}_m\leq 2$ with $m=2,\ldots,N$ are the real eigenvalues of $\mathcal{L}^{(\gamma)}$. Processes where $p_0^{(\gamma)}(t)$ decays rapidly to the $\lim_{t\to \infty} p_0^{(\gamma)} (t)=N^{-1}$ explore efficiently new sites of the network.  

To analyze in detail this average return probability, let us use as before the case of a ring. By using the spectrum of the Laplacian matrix of a ring and the definition of $\mathcal{L}^{(\gamma)}$, we can determine analytically the spectrum $\{\zeta_l ^{(\gamma)}  \}_{l=2}^N$ for the modified fractional Laplacian. With these eigenvalues and Eq. (\ref{Pretfrac}), in the limit $N\to\infty$, and approximating the sum by an integral as before, we obtain:
\begin{equation}\label{PretLine}
p_0^{(\gamma)}(t)=\frac{1}{2\pi}\int_0^{2\pi}e^{-(2-2\cos\theta)^{\gamma} t/k^{(\gamma)}}d\theta .
\end{equation}
Here $k^{(\gamma)}=-\frac{\Gamma(-\gamma)\Gamma(2\gamma+1)\sin(\pi \gamma)}{\pi\Gamma(1+\gamma)}$ denotes the elements in the diagonal of $\mathbf{L}^{\gamma}$ given by Eq. (\ref{LfracLine}). In the limit $\gamma=1$ we recover the well-known analytical result for normal random walks where $p_0^{(1)}(t)= e^{-t} I_0(t)$, where $I_n(x)$ denotes the modified Bessel function of the first kind \cite{RednerBook}. Not only for $\gamma=1$, but even for some rational values of $\gamma$, we can obtain an analytical result for this quantity. For example, when $\gamma=1/2$ we have $p_0^{(1/2)}(t)= I_0\left(\frac{\pi  t}{2}\right)-\text{L}_0\left(\frac{\pi  t}{2}\right)$, where $\text{L}_n(x)$ denotes the modified Struve function \cite{HandbookAbramowitz}. On the other hand, for $t\gg1$, Eq. (\ref {PretLine}) can be expressed as:
\begin{equation}
p_0^{(\gamma)}(t)\sim\frac{1}{\pi}\int_0^{\pi}\exp\left[\frac{-2^{2\gamma}\theta^{2\gamma}}{k^{(\gamma)}} t\right]d\theta,
\end{equation}
which can be evaluated analytically and takes the form \cite{HandbookAbramowitz}:
\begin{equation}
p_0^{(\gamma)}(t)\sim\frac{(k^{(\gamma)})^{\frac{1}{2 \gamma}}}{4\pi\gamma}  \left[\Gamma\left(\frac{1}{2\gamma }\right)-\Gamma\left(\frac{1}{2\gamma},\nu t\right)\right] t^{-\frac{1}{2 \gamma}},
\end{equation}
where $\nu=(2\pi)^{2 \gamma}/k^{(\gamma)}$ and $\Gamma(a,x)$  is the incomplete Gamma function \cite{HandbookAbramowitz}. Therefore, the average return probability $p_0^{(\gamma)}(t)$ for a ring, with $N\to\infty$ in the limit $t\gg1$, decays as a power-law  given by $p_0^{(\gamma)}(t)\sim t^{-1/(2\gamma)}$. This result generalizes the case $\gamma=1$ for a normal random walk, where $p_0^{(1)}(t)\sim t^{-1/2}$ \cite{RednerBook,VespiBook}.
\begin{figure}[!t] 
\begin{center}
\includegraphics*[width=0.48\textwidth]{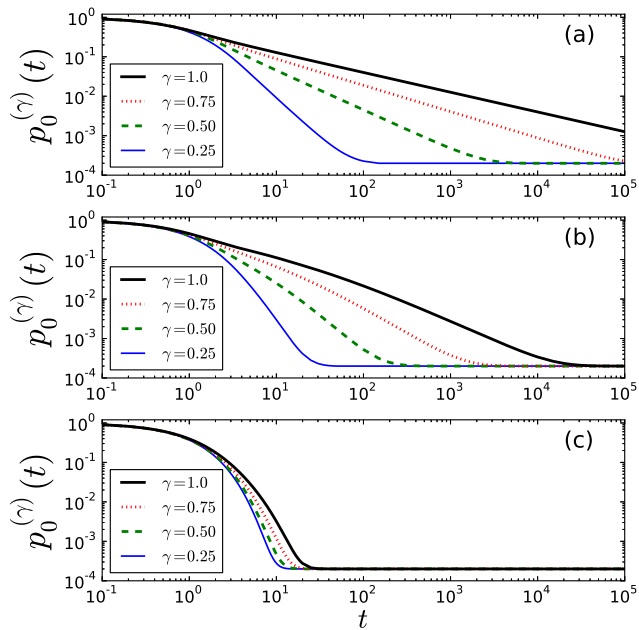}
\end{center}
\vspace{-5mm}
\caption{\label{Fig3}(Color online) Average fractional return probability $ p_0^{(\gamma)} (t)$ as a function of time for different networks with $N=5000$ nodes and calculated using Eq. (\ref{Pretfrac}). (a) A ring,  (b) a tree, and (c) a scale-free (SF) network of the Barab\'asi-Albert type \cite{BarabasiAlbert}. We used different values of $\gamma$ as indicated in the insets.} 
\end{figure}
\begin{figure}[!t]
\begin{center}
\includegraphics*[width=0.48\textwidth]{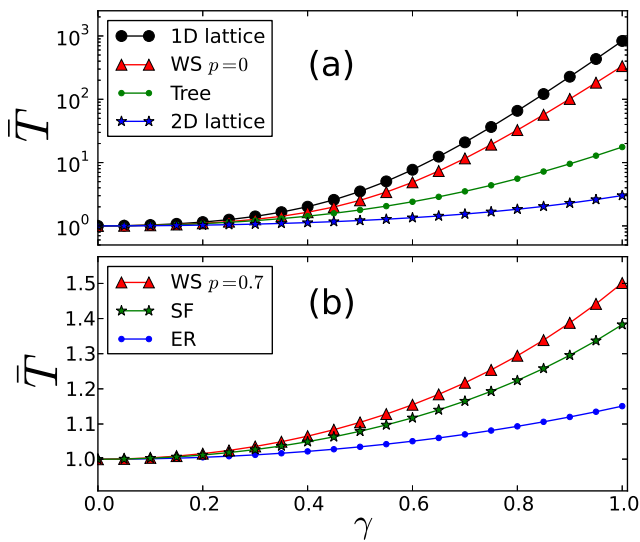}
\end{center}
\vspace{-5mm}
\caption{\label{Fig4} (Color online) Average time $\bar{T}$ vs $\gamma$ calculated from Eq. (\ref{globalT}) for different types of networks with $N=5000$ nodes. (a) Large-world networks: ring (1D lattice), the regular network with degree $k=4$ obtained by the Watts-Strogatz (WS) model with rewiring probability $p=0$ \cite{WattsStrogatz}, a tree, a 2D lattice with dimension $50\times 100$ and periodic boundary conditions. (b) Small-world networks: the WS network used in (a) with rewiring probability $p=0.7$, a scale-free (SF) network of the Barab\'asi-Albert type \cite{BarabasiAlbert}, and Erd\H{o}s-R\'enyi (ER) network with $p=\log N/N$ \cite{ErdosRenyi}.}
\end{figure}

Let us analyze now this quantity for general networks in more detail. In Fig. \ref{Fig3} we show the average fractional return probability $p_0^{(\gamma)}(t)$ using the eigenvalues of the modified fractional Laplacian $\mathcal{L}^{(\gamma)}$ and Eq. (\ref{Pretfrac}). Notice that the fractional dynamics explores more efficiently the networks in comparison with the normal case $\gamma=1$ (i.e. $p_0^{(\gamma)}(t)$ decays more rapidly to the asymptotic value $N^{-1}$). In particular, in Fig. \ref{Fig3}(a) we observe the power-law decay of $p_0^{(\gamma)}\propto t^{-1/(2\gamma)} $, as predicted by our analytical results for a ring, before reaching the value $N^{-1}$. The effect of the efficiency of navigation due to fractional dynamics is more noticeable for large-world networks (ring and tree) than for small-world (scale-free) networks. 

In order to quantify the efficiency to explore the network, we introduce a global time $\bar{T}$ as:
\begin{equation}\label{globalT}
\bar{T}\equiv \int_0^\infty(p_0^{(\gamma)}(t)-p_0^{(\gamma)}(\infty))dt=\frac{1}{N}\sum_{j=2}^N \frac{1}{\zeta_j^{(\gamma)}}.
\end{equation}
The inverse of the eigenvalues $\zeta_j^{(\gamma)}$ are the characteristic times $\text{T}_j^{(\gamma)}$ that dominates the dynamics; these times are also relevant for the problem of synchronization in networks \cite{ArenasPhysRep2008}. Thus, this global time $\bar{T}$ is the average of these characteristic times in $p_0^{(\gamma)}(t)$. In the context of Markovian processes $\tau\equiv N \bar{T}$ is the Kemeny's constant, that for random walks is the global time $\tau=\sum_{j\neq i}^N p_j^{\infty}\langle T_{ij} \rangle$, where $\langle T_{ij}\rangle$ is the mean first passage time (MFPT) defined as the mean number of steps taken to reach the node $j$ for the first time starting from the node $i$ \cite{Hughes,RiascosMateos,Kemeny,Zhang}. 

In Fig. \ref{Fig4} we show the global time $\bar{T}$ as a function of $\gamma$ for different types of networks. In Fig. \ref{Fig4}(a), for large-world networks, the effect  of the fractional dynamics reduces several orders of magnitude the time $\bar{T}$, in comparison with the case $\gamma=1$.  In Fig. \ref{Fig4}(b) we show that even for small-world networks the fractional dynamics improves the efficiency to explore the networks. 
\section{Conclusions}
In summary, we have introduced a formalism of fractional diffusion on networks based on a fractional Laplacian matrix that can be constructed from the spectra and eigenvectors of the Laplacian matrix. This fractional approach allows random walks with long-range dynamics providing a general framework for anomalous diffusion and navigation in networks. We obtained exact results for the stationary probability distribution, the average fractional return probability and a global time. Based on these quantities, we show that the efficiency to navigate the network is greater if we use a fractional random walk, in comparison to a normal random walk. For the case of a ring, we obtain exact analytical results showing that the fractional transition and return probabilities follow a long-range power-law decay and, thus, the emergence of L\'evy flights on networks. Our general fractional diffusion formalism applies to regular, random, and complex networks and can be implemented from the spectral properties of the Laplacian matrix, providing an important tool to analyze anomalous diffusion on networks. Our results show how the long-range displacements improve the efficiency to reach any node of the network inducing dynamically the small-world property in any structure.
\section{Acknowledgments}
A.P.R. acknowledges support from CONACYT M\'exico.

\providecommand{\noopsort}[1]{}\providecommand{\singleletter}[1]{#1}%
\end{document}